\documentclass{PoS}

\title{Constraining Sea Quark Distributions Through $W^{\pm}$ Cross Section Ratios Measured at STAR}

\ShortTitle{$W^{\pm}$ Cross Section Ratios Measured at STAR}

\author{\speaker{M.~Posik} (for the STAR Collaboration)\\
        Temple University, Philadelphia, PA, USA\\
        E-mail: \email{posik@temple.edu}}

\abstract{
Over the past several years the STAR experiment at RHIC has been contributing to our understanding of the proton structure. Through its instrumentation, STAR is well equipped to measure $W \rightarrow \nu + e$ in $\sqrt{s}$ = 500/510 GeV proton-proton collisions  at mid-rapidity (-1.1 $\le \eta \le $ 1.1) . The $W$ cross section ratio ($W^+/W^-$) is sensitive to unpolarized $u$, $d$, $\bar{u}$, and $\bar{d}$ quark distributions. At these kinematics, STAR is able to measure the quark distributions near Bjorken-$x$ values of 0.1. The RHIC runs in 2011, 2012 and 2013 at $\sqrt{s}$ = 500/510 GeV saw a significant increase in delivered luminosity from previous years. This resulted in a total data sample being collected of about 352 pb$^{-1}$ of integrated luminosity. The increased statistics will lead to a higher precision measurement of the $W^+/W^-$ cross section ratio than was previously measured by STAR's 2009 run,  as well as allow for a measurement of its $\eta$ dependence at mid-rapidity. Presented here is an update of the $W$ cross section ratio analysis from the STAR 2011, 2012 and 2013 runs. 
}

\FullConference{The XXIII International Workshop on Deep Inelastic Scattering and Related Subjects\\
		April 27 - May 1, 2015\\
                Southern Methodist University\\
		Dallas, Texas 75275}

\begin{document}

\section{Motivation}
Over the past several years parton distribution functions (PDFs) have been becoming more and more precise~\cite{CT10Ana,HERAFitter,NNPDF}. However, there are still regions in which more precision data is needed which can be used to help constrain the PDFs. For example the sea quark distributions near the valence region, $x$ $\sim$ 0.1-0.3, still have sizable uncertainties~\cite{CT10}. 

One of the data sets used to determine the anti-quark PDFs is the $\bar{d}/\bar{u}$ measurement from E866~\cite{E866}, which measured $\bar{d}/\bar{u}$ to good precision at lower $x$ ($x < 0.15$). However, their precision quickly deteriorates as they approach higher $x$ ($x>$ 0.2). These data suggest an interesting behavior, as $x$ increases there seems to be a transition from being $\bar{d}$ dominated to $\bar{u}$ dominated around $x\sim$ 0.25. Many models are able to describe the general $\bar{d} > \bar{u}$ behavior seen at low $x$, but fail to predict the suggested $\bar{u} > \bar{d}$ transition~\cite{Peng:2011}. To better determine the behavior of $\bar{d}/\bar{u}$ the experiment SeaQuest (E-906)~\cite{E906} has been designed and is currently running. Through Drell-Yan scattering, SeaQuest will probe the sea quark distribution at lower $Q^2$ than E866, but increase the precision and $x$ reach of the $\bar{d}/\bar{u}$ measurement.  Although this will help constrain the PDF fits, ideally one would like more data to fit from different scattering processes and $Q^2$ scales. This will help to add more independent data to global fits, and serve as a cross check of our understanding of the QCD sea.

The $W$ boson production in proton-proton collisions is also sensitive to the sea quarks. The $W^+$ boson is sensitive to the $\bar{d}$ quark, while the $W^-$ boson is sensitive to the $\bar{u}$ quarks which can be seen in equation~\ref{eq:quark}, and probes the distribution at $Q^2 \sim M_{W}^{2}$. The leptonic decay from $W$ bosons can be detected by looking for leptons with a high transverse momentum, $p_T$, near $M_{W}/2$. Then a charge separation of the leptons can be used to determine which charged $W$ boson they decayed from.   

\begin{equation}\label{eq:quark}
  u + \bar{d} \rightarrow W^+ \rightarrow e^+ + \nu, \;\; d + \bar{u} \rightarrow W^- \rightarrow e^- + \bar{\nu}.
\end{equation}     

By considering the leading order expression for the charged $W$ cross section ratio~\cite{Soffer94}, $\frac{\sigma_{W^+} }{\sigma_{W^-}}$ ($R_W$), the direct relationship to the sea quarks can be seen 

\begin{equation}\label{eq:RW}
  R_W \equiv \frac{ \sigma_{W^+} }{ \sigma_{W^-} } \sim \frac{u\left(x_1\right)\bar{d}\left(x_2\right) + \bar{d}\left(x_1\right)u\left(x_2\right)}{\bar{u}\left(x_1\right)d\left(x_2\right) + d\left(x_1\right)\bar{u}\left(x_2\right)}.
\end{equation}

\noindent It should be noted that although $R_W$ can be measured at the LHC, the region of $x$ that would be probed is below the valence region near $x\sim 0.08$ (assuming a $\sqrt{s}$ = 1 TeV and $\eta$ = 0).

\section{Experiment}

The STAR experiment at RHIC~\cite{STAR} serves as an excellent place to measure the charged $W$ cross section ratio, which was first measured in the STAR 2009 run~\cite{STAR2012}. The STAR experiment measured $R_W$ using proton-proton collisions at center of mass energies of $\sqrt{s} = 500/510$ GeV in the mid-rapidity region ($-1.1 \le \eta \le 1.1$). Several sub-detectors were used to select the $W$ events and separate their charge: the time projection chamber (TPC)~\cite{TPC}, used for particle tracking, and the barrel electromagnetic calorimeter (BEMC)~\cite{BEMC}, used to measure particle energy. A third sub-detector, the endcap electromagnetic calorimeter (EEMC)~\cite{EEMC}, was used to estimate the background contributions. The mid-rapidity region of STAR corresponds to about $0.1 \le x \le 0.3$ and $Q^{2} \sim M_{W}^2$, which could have an impact on constraining PDFs as this is the $x$ region where E866's precision starts to drop off and is the region where the data suggests that the $\bar{u}$ quark density is greater than the  $\bar{d}$ quark density. STAR has taken advantage of the yearly increase in luminosity that RHIC has provided. This luminosity increase has led to roughly 352 pb$^{-1}$ of integrated luminosity being collected during the 2011-2013 runs. With the 2013 data set still under analysis, a preliminary $R_W$ result is presented using only a fraction (102 pb$^{-1}$) of the collected 2011-2013 data. 

\section{Results}
\begin{figure}[!h]
\centering
\includegraphics[width=0.75\columnwidth]{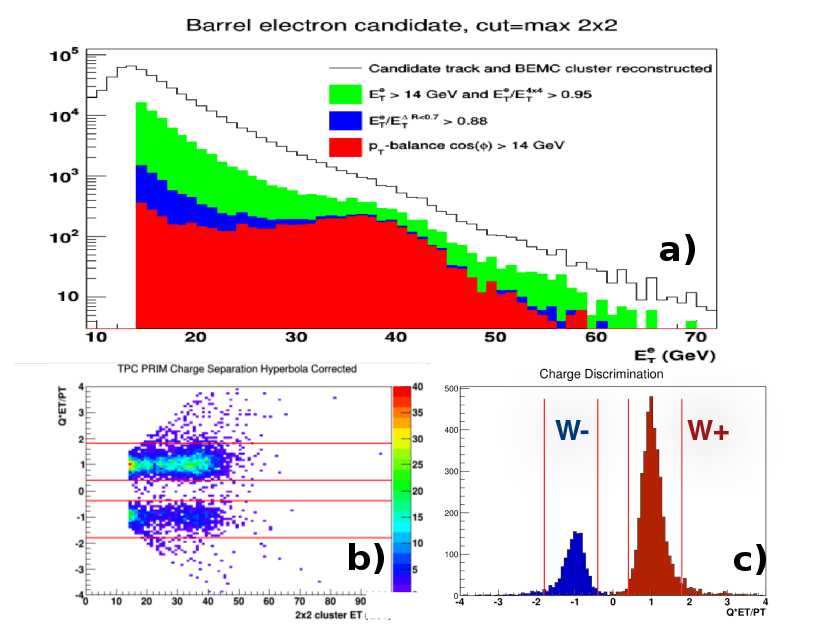}
\caption{Some cuts applied to data used to select leptons which likely originated from $W$ decay. a) Application of several isolation cuts including a minimum $E_T$ cut, electron energy ratio cuts, and a signed $p_T$ cut.  b) Charge separation cut vs. $E_{T}$.  c) Projection of the charge separation vs. $E_T$ projected onto the charge separation axis.}
\label{fig:cuts}
\end{figure}

The leptons from $W$ decay are selected by following the methodology previously established by STAR~\cite{STAR2012}. Several cuts which include matching high $p_T$ tracks to BEMC clusters, a series of isolation cuts used to isolate the leptons, a $p_T$-balance cut which looks for the large missing neutrino momentum, and a charge separation cut are applied to select leptons that are likely produced from $W$ decay. Figure~\ref{fig:cuts} shows the application of several isolation cuts and charge separation cuts to the data. In panel a), one can see that as more isolation cuts are applied there is a decrease in background events, which populate the kinematic region $E_T < 25$ GeV, and an enhancement of the lepton signal near $E_T \sim M_W/2$. Panels b) and c) show cuts applied to the data in order to select events which have likely originated from $W^{+}$ or $W^{-}$ decays. Panel b) shows the charge separation as a function of $E_T$, while panel c) projects the charge separation as a function of $E_T$ on to the charge separation axis. The charge separation cuts are indicated by the red lines and were chosen to avoid contamination from the opposite charge.  

The charged $W$ cross section ratio can be measured experimentally as

\begin{equation}\label{eq:Exp-RW}
  \frac{\sigma_{W^+}}{\sigma_{W^-}} = \frac{\left(N^+_O - N^+_B\right)}{\left(N^-_O - N^-_B\right)}\frac{\epsilon^-}{\epsilon+},
\end{equation}

\noindent where $\pm$ corresponds to positively or negatively charged lepton, $N_O$ are the number of events that pass the lepton selection cuts, $N_B$ are the number of background events estimated to be contaminating the data set, and $\epsilon$ is the efficiency at which $W$ events are detected. 

\begin{figure}[!h]
\centering
\includegraphics[width=0.75\columnwidth]{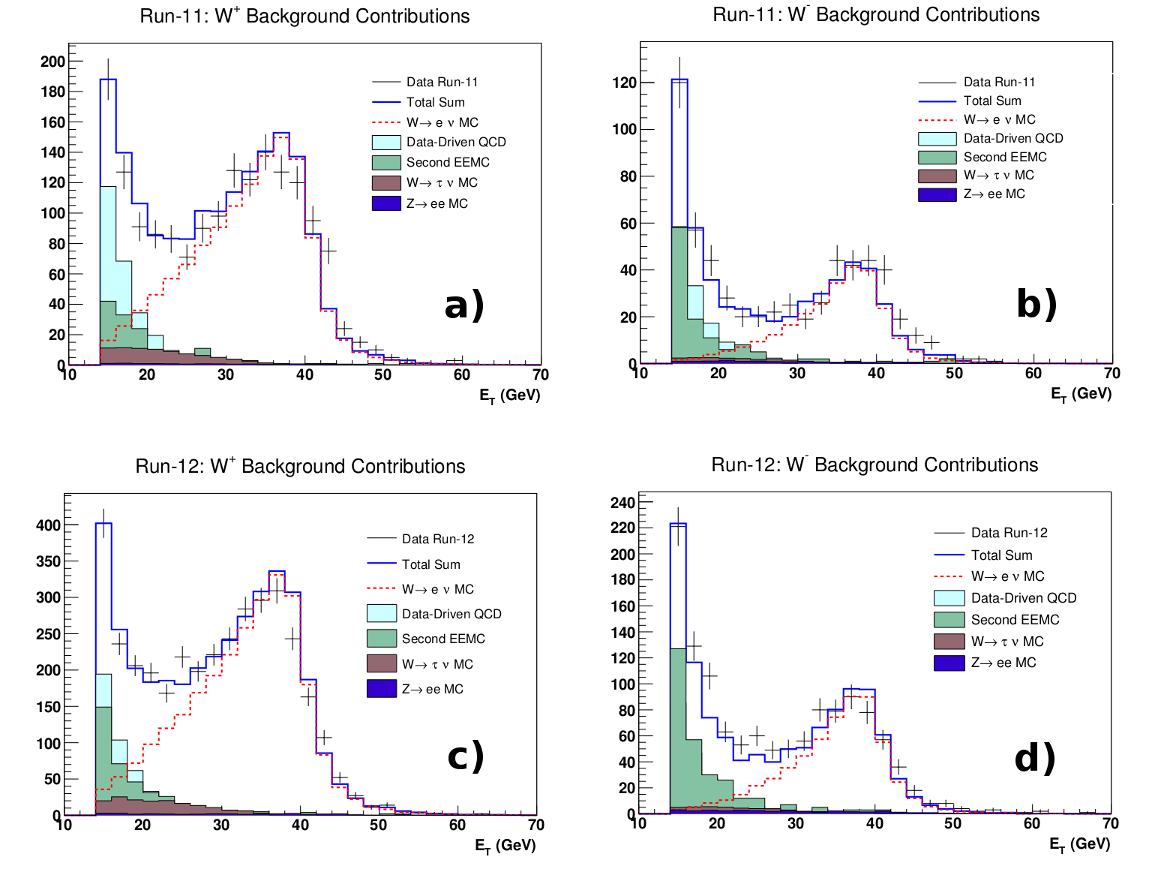}
\caption{Background and Monte Carlo contributions compared to data. a) Run-11 $W^{+}$, b) Run-11 $W^{-}$, c) Run-12 $W^{+}$, and d) Run-12 $W^{-}$. }
\label{fig:background}
\end{figure}

Figure~\ref{fig:background} shows the various background contributions, Monte Carlo simulation of the $W$ decay (based on Pythia 6.4.22~\cite{Pythia} and GEANT~\cite{GEANT}), and a comparison of the data to Monte Carlo $W$ signal with background contributions included for the 2011 and 2012 data sets. The background contribution labeled {\it{Second End Cap}} is an estimate of the background caused by an escaping jet's $p_T$ being misidentified as the neutrino's missing $p_T$. This is predominately a QCD like background. When the final cut of $E_T > 25$ GeV is applied, there is very little background contributions from $W\rightarrow \tau + \nu$ and $Z\rightarrow ee$ decays. The background was found to be dominated by QCD background.

\begin{figure}[!h]
\centering
\includegraphics[width=0.75\columnwidth]{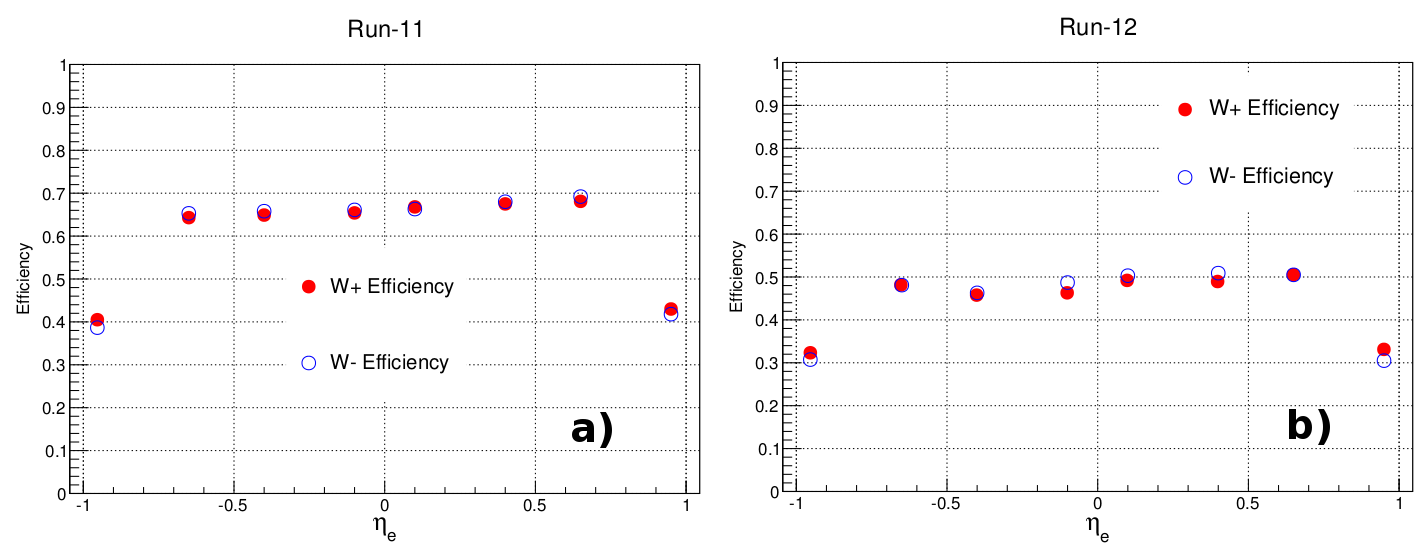}
\caption{$W^{+}$ and $W^{-}$ efficiencies as a function of electron pseudo-rapidity for a) Run-11 and b) Run-12. }
\label{fig:eff}
\end{figure}

A Monte Carlo based on Pythia 6.4.22~\cite{Pythia} and GEANT~\cite{GEANT} is used to determine the $W^\pm$ detection efficiencies, shown in Fig.~\ref{fig:eff}. These efficiencies account for all cut and detector efficiencies. The 2011 data was found to have a higher efficiency than the 2012 data due to running at a higher luminosity rate in 2012. Running at a higher instantaneous luminosity lead to more pile-up in the TPC, which resulted in less efficient track reconstruction and hence less efficient $W$ detection. However in both data sets there was only a small ($\sim$ 1-2\%) charge dependence measured between the $W^+$ and $W^-$ efficiencies, which means the $\frac{\epsilon^-}{\epsilon^+}$ factor will have a negligible contribution to the charged $W$ cross section ratio.
 
\begin{figure}[!h]
\centering
\includegraphics[width=0.5\columnwidth]{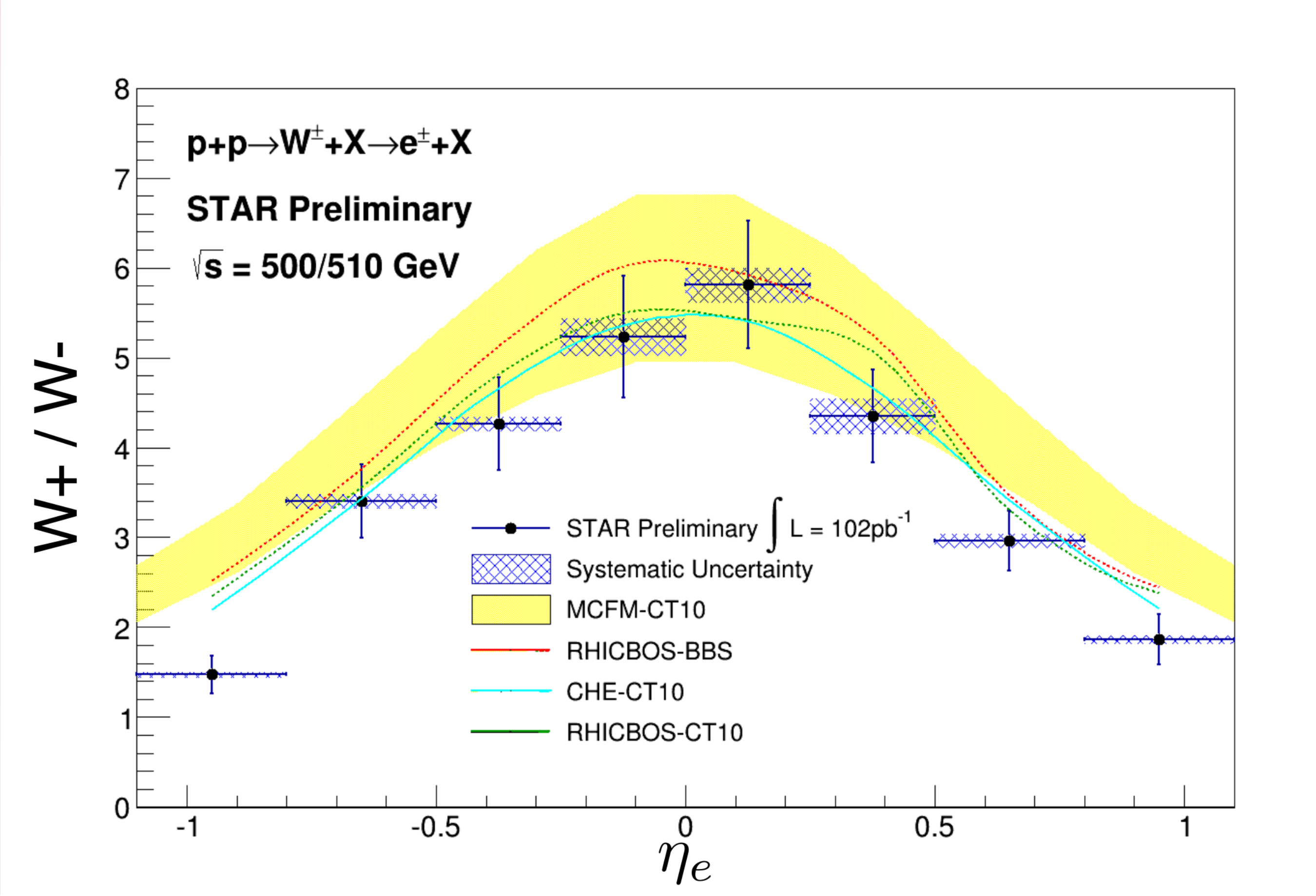}
\caption{$W^+$/$W^{-}$ cross section ratio as a function of electron pseudo-rapidity.}
\label{fig:RW-electron}
\end{figure}

\begin{figure}[!h]
\centering
\includegraphics[width=0.5\columnwidth]{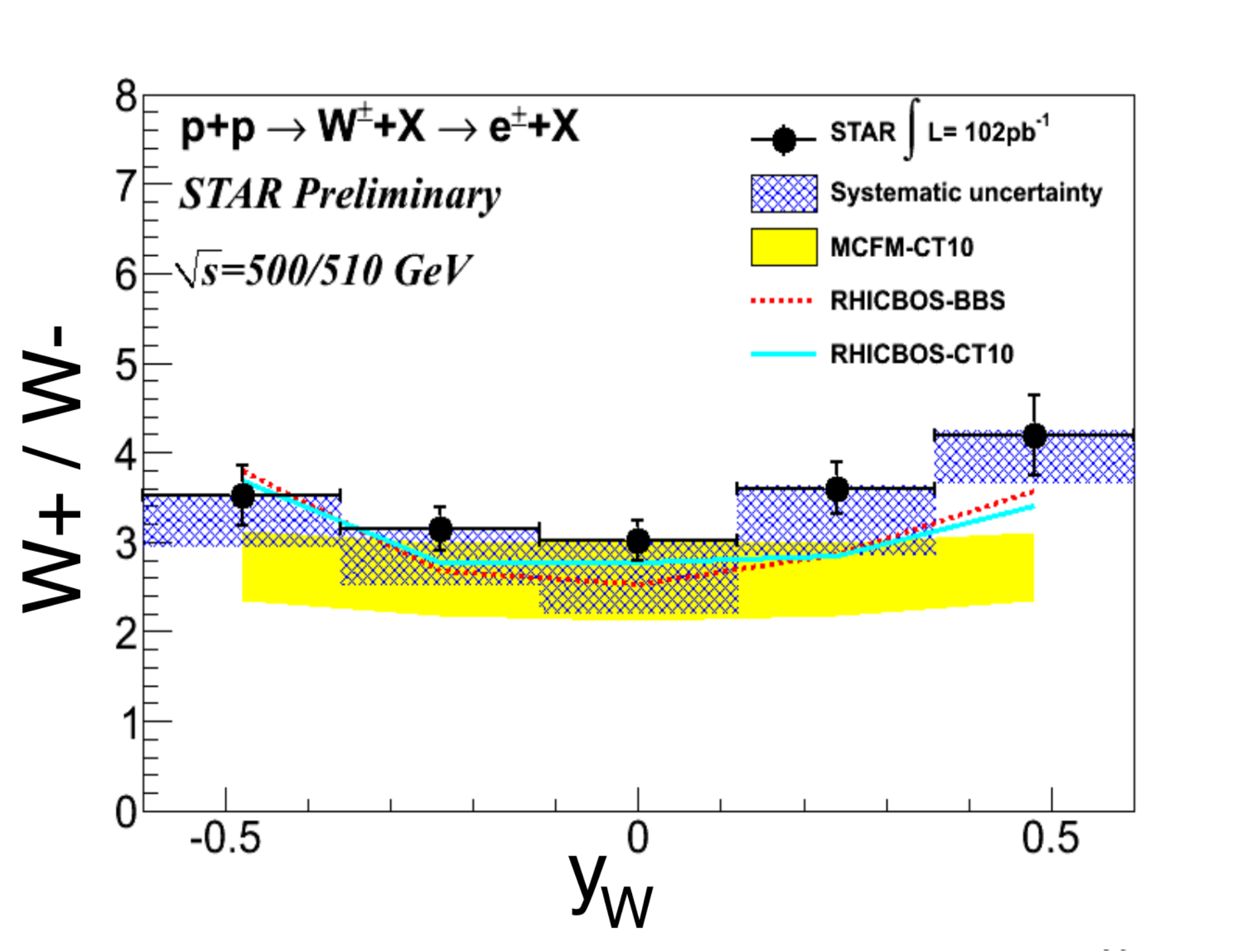}
\caption{$W^+$/$W^{-}$ cross section ratio as a function of the $W$ boson rapidity.}
\label{fig:RW-boson}
\end{figure}

Figure~\ref{fig:RW-electron}(\ref{fig:RW-boson}) shows the charged $W$ cross section ratio for the combined 2011 and 2012 runs, computed using equation~\ref{eq:Exp-RW}, as a function of the electron pseudo-rapidity, $\eta_e$ ($W$ boson rapidity, $y_W$). More information on how the $W$ boson kinematics were reconstructed can be found in~\cite{Sal14,Sal15}. The error bar on the data points represents the statistical uncertainty, while the shaded boxes correspond to the systematic uncertainty. The yellow band and colored curves serve as a comparison to different PDF sets~\cite{CT10nlo,BBS} and theory frame works~\cite{MCFM,RHICBOS01}. Note that the systematic uncertainties for the charged $W$ cross section ratios as a function of $\eta_e$ are well under control and we are dominated by our statistical precision. Further studies into the newly established $W$ boson reconstruction process~\cite{Sal14,Sal15} should reduce the systematic uncertainties on the $W^{\pm}$   cross-section ratio dependence on the boson kinematics.

\section{Summary}
We have measured and presented charged $W$ cross section ratios from combined 2011 and 2012 proton-proton STAR data at $\sqrt{s} = 500/510$ GeV. The inclusion of this data into global PDF analysis should help constrain the sea quark distributions and provide additional insight into the $\bar{d}/\bar{u}$ ratio near the valance region. Furthermore, with the inclusion of the STAR 2013 data ($\sim$ 250 pb$^{-1}$), we will be able to further improve on the precision of our charged $W$ cross section ratios.

\end{document}